\documentclass[reprint, superscriptaddress]{revtex4-2}

\usepackage[utf8x]{inputenc}

\usepackage{amsmath}
\usepackage{amsfonts} 
\usepackage{graphicx}
\graphicspath{ {./} }
\usepackage{afterpage}
\usepackage[section]{placeins}
\usepackage[hidelinks]{hyperref}
\usepackage{xcolor}
\usepackage{siunitx}
\usepackage{mathtools}
\usepackage{layouts}
\usepackage{booktabs}
\usepackage{ulem}
\usepackage{enumitem}
\usepackage{tabularx}
\usepackage{upgreek}
\usepackage[caption=false]{subfig}
\usepackage{pdfcomment}

\newcommand{\Imperial}{QOLS, Blackett Laboratory, Imperial College London, South Kensington, London SW7 2AZ, United Kingdom}
\newcommand{\Toshiba}{Toshiba Europe, Ltd., 208 Cambridge Science Park, Milton Road, Cambridge, CB4 0GZ, United Kingdom}

\newcommand{\inj}{_\textrm{inj}}

\begin{document}

\title{Quantified Effects of the Laser Seeding Attack in Quantum Key Distribution}

\date{\today}

\author{V. Lovic}
\affiliation{\Toshiba}
\affiliation{\Imperial}
\author{D. G. Marangon}
\email{Corresponding author: davidegiacomo.marangon@unipd.it\\Current address: Department of Engineering of Information, University of Padova, Italy}
\author{P. R. Smith}
\email{Corresponding author: raymond.smith@toshiba.eu}
\author{R. I. Woodward}
\author{A. J. Shields}
\affiliation{\Toshiba}

\begin{abstract}
Quantum key distribution (QKD) enables private communications with information-theoretic security. To guarantee the practical security of QKD, it is essential that QKD systems are implemented in accordance to theoretical requirements and robust against side-channel attacks. Here we study a prominent attack on QKD transmitters known as the laser seeding attack (LSA). It consists in injecting photons into the laser of the transmitter in an attempt to modify the outgoing light in some way that is beneficial to the eavesdropper. In this work we measure the response of a QKD transmitter to the LSA as a function of the optical power injected, allowing us to quantify the level of optical attenuation required to mitigate the attack. Further, we employ a laser rate equation model to numerically simulate the effects of the LSA on a gain-switched laser. With this model we are able to reproduce previous experimental results, as well as generate new insight into the LSA by examining the effects of the LSA when the QKD transmitter is operated with different laser current driving parameters. 
\end{abstract}

\maketitle

\section{Introduction}
Quantum key distribution (QKD) is a mature quantum technology that can be used to establish a secret key between two communicating parties, conventionally referred to as “Alice” and “Bob” \cite{bennett_quantum_1984, gisin_quantum_2002, pirandola_advances_2020}.
QKD
enables communications with information theoretic security and has therefore attracted great interest in the face of the threat posed by quantum computers to public-key cryptography \cite{shor_polynomial-time_1997}.
The security of QKD rests on mathematical proofs that make certain assumptions about the physical systems that implement the protocol \cite{shor_simple_2000, gottesman_security_2004-1}. 
It is therefore crucial that these physical systems conform to the assumptions made by the theory.
During the last two decades of QKD research there has been a two-sided effort to bridge this gap between theory and practice: on the one hand QKD security proofs and protocols have advanced, making fewer and more realistic assumptions \cite{gottesman_security_2004-1, lo_measurement-device-independent_2012, pereira_quantum_2019}; on the other, physical QKD systems have moved closer to theoretical requirements by implementing countermeasures to known security vulnerabilities \cite{xu_secure_2020, lucamarini_practical_2015, yuan_resilience_2011, qian_robust_2019}.

\begin{figure}
	\includegraphics[width=\linewidth]{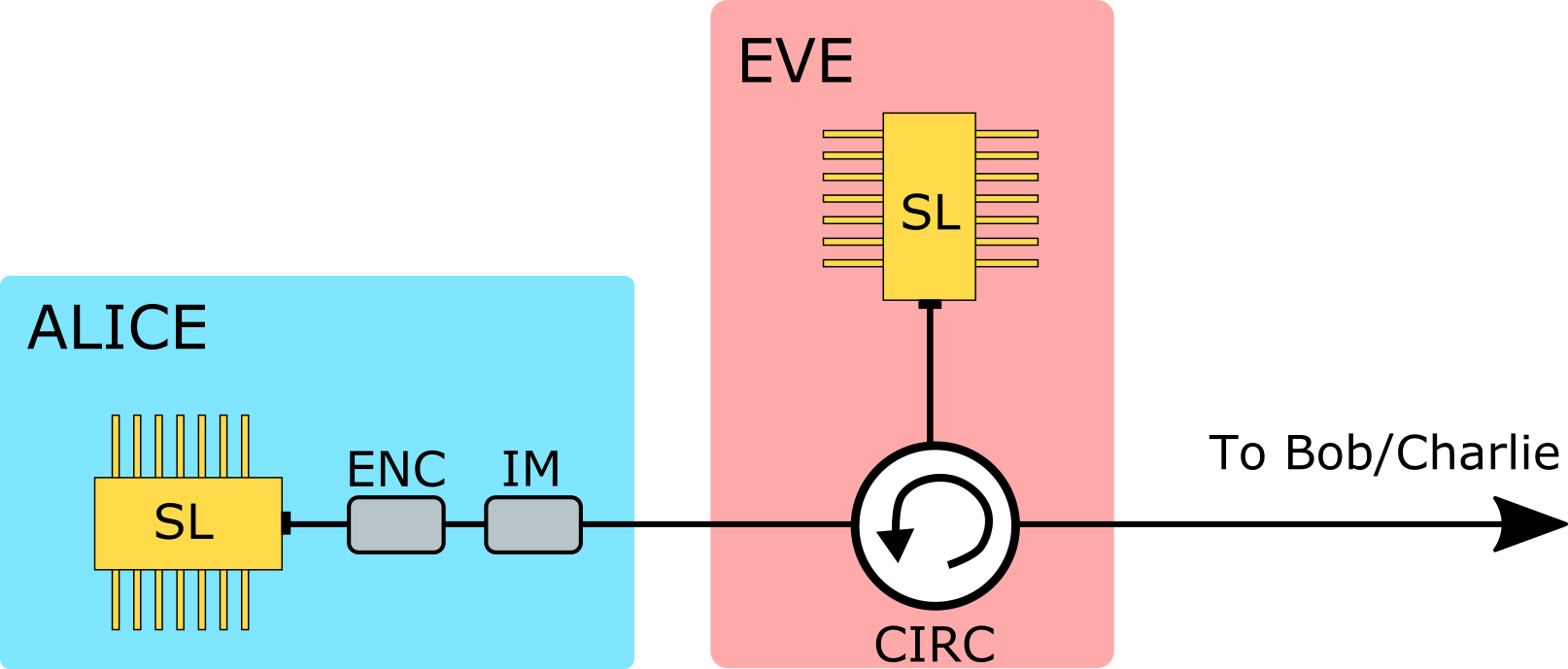}
	\caption{\label{fig:simple_LSA}A simplified QKD transmitter (Alice), with a simple implementation of the LSA by Eve. Eve can add a circulator (CIRC) to the quantum channel between Alice and Bob such that she can inject light into Alice's transmitter while letting Alice's light pass through unimpeded. ; ENC, encoder; IM, intensity modulator; SL, semiconductor laser.}
\end{figure}

QKD implementation security vulnerabilities can be broadly classified as targeting either the transmitter or the receiver. Historically, the most serious proposed attacks against QKD systems have targeted the receiver. However, the development of measurement-device-independent QKD and its variants has provided a solution to all known and possible vulnerabilities in the receiver \cite{braunstein_side-channel-free_2012, lo_measurement-device-independent_2012, lucamarini_overcoming_2018}. The focus of QKD implementation security has therefore shifted toward the transmitter \cite{xu_secure_2020}. 
In this work we study a prominent attack on QKD transmitters known as the laser seeding attack (LSA) \cite{sun_effect_2015, lee_free-space_2017, pang_hacking_2020, huang_laser-seeding_2019, zhang_analysis_2022}. 
This attack consists of an eavesdropper, named “Eve”, injecting light into the laser of a QKD transmitter to try to change the emitted light in some way that is beneficial to her (Fig.~\ref{fig:simple_LSA}). For example, it was demonstrated in Ref.~\cite{sun_effect_2015}  how the LSA can violate the assumption of a phase-randomized source: by seeding the QKD transmitter laser with light of a known phase, the LSA can give an eavesdropper full knowledge of the phase of the outgoing light, which drastically reduces the performance of a QKD system. In Ref.~\cite{sun_effect_2015} the LSA was implemented with use of only two different levels of injected power. Additionally, the LSA was implemented with use of an isolated laser diode. The isolation of an isolated laser diode is difficult to characterize, and so it is unclear exactly how much light was reaching Alice's laser cavity. Altogether, this means that we do not know accurately how the effects of the LSA depend the level of injected power (that reaches the laser cavity). This is a crucial consideration for Alice when she is implementing countermeasures to the attack: if Eve must inject a large amount of light for a successful attack, then the attack will be easier to detect. However, if Eve can implement a successful LSA with only a small amount of injected power, it will be more difficult to detect or prevent. In the first part of this work (Section \ref{sec:phase}), we implement the LSA experimentally and measure its effects on the phase randomization of Alice's laser as a function of the injected optical power. We accurately measure the level of injected power that reaches Alice's laser cavity by using an unisolated laser, and measuring the incoming light with a power meter. We find that Eve can influence the phase of Alice's laser using much lower levels of injected power than may have been considered previously \cite{sun_effect_2015, huang_laser-seeding_2019}. Additionally, by accurately measuring how the effects of the LSA vary with increasing injected power, we are able to quantify the level of optical isolation required to limit the effects of the LSA on Alice's laser. This is of obvious practical relevance to the secure implementation of QKD systems.

In Ref.~\cite{sun_effect_2015} it is also suggested that the LSA could have other damaging effects apart from derandomizing the phase. Indeed, it is demonstrated in Ref.~\cite{huang_laser-seeding_2019}  that the LSA can also increase the power output of a QKD transmitter, and the effect of this increase on the secure key rate is quantified. However, there are still other effects of the LSA, such as a shift in the wavelength of Alice's laser \cite{lee_free-space_2017, zhang_analysis_2022}, a reduced turn-on delay, or a change in the shape of the emitted pulse, and Eve can try to use any of these to her advantage. In the second part of this work (Sections \ref{sec:LSA_OIL} and \ref{sec:intensity}) we use a laser rate equation model that allows us to simulate the output of a laser subjected to the LSA by modeling it as a form of optical injection locking (OIL). We argue that the laser rate equations are a useful tool for the study of the LSA and can be used as a general model of the LSA to investigate all of its various effects. To demonstrate this usefulness, we use the model to reproduce, in simulation, previous published experimental results in the literature \cite{sun_effect_2015, huang_laser-seeding_2019}. Further, we use the model to generate new insight into the effects of the LSA on QKD transmitters by examining how its effects on the power output of a gain-switched laser vary under different laser current driving conditions.

\section{Effects on Phase Randomization} \label{sec:phase}
QKD has been proven to be secure with and without phase randomized signals \cite{gottesman_security_2004-1, lo_security_2007-1}. However, the use of nonrandom phase leads to a significantly worse performance. Hence, current implementations of QKD use phase randomized pulses of light. The phase randomization can be achieved with the use of a phase modulator connected to a cryptographically secure random number generator, but this adds cost and complexity to a QKD system. A widely used, simple, and effective alternative to adding a phase modulator is to use a gain-switched laser diode to generate the pulses of light \cite{paraiso_advanced_2021}. Gain-switching consists in driving a laser diode alternately above and below its lasing threshold. When the laser is below threshold, the cavity empties such that when the laser is next driven above threshold, the pulse that is generated has a random phase from spontaneously emitted photons in the cavity \cite{jofre_true_2011}. Gain-switched lasers can produce short, naturally phase randomized pulses of light at gigahertz clock rates  by adjustment of onlu the current signal driving the laser and are therefore widely used in modern QKD implementations \cite{paraiso_advanced_2021}. We therefore assume that the target of the LSA is a gain-switched laser diode, as widely used in current real-world QKD systems \cite{yuan_10-mb/s_2018, boaron_secure_2018}.

The current parameters of a gain-switched laser diode must be set carefully to ensure that the laser cavity fully empties between each pulse. If photons from a previous pulse are still present in the cavity when the laser is driven above threshold, then the new pulse will inherit its phase from these photons This will therefore lead to correlations between the phase of the emitted pulses, which is detrimental to the security of QKD. 

The LSA works in a similar way: by injecting external photons into the laser cavity the generated pulses will inherit their phase from the external photons, rather than from spontaneous emission, preventing the phase from being randomized between pulses. Additionally, the attacker can deterministically control the phase of the injected photons, and therefore of the emitted pulses, which is a further security concern. 

\subsection{Experiment}
A simple implementation of the LSA consists in Eve injecting constant, continuous wave laser light into Alice's laser. As explained, this prevents the phase randomization in Alice's laser, locking its phase to a constant value determined by the coherent injected light. 

Alice may try to detect this attack by monitoring the phase at the output of her laser to check for signs of nonrandomness. However, with a successful LSA, Eve can deterministically control the phase of Alice's laser. Therefore, we propose and experimentally demonstrate a  more sophisticated version of the LSA, which we refer to as the ``phase-randomized LSA''. In this version of the LSA, Eve injects light with a seemingly random phase, but which is nonetheless completely known to her, into Alice's transmitter. For example, she can modulate the phase of her light with a phase modulator connected to a pseudo random number generator under her control, or alternatively she can gain-switch her laser and measure the phase before it is sent into Alice's transmitter.  Either way, when Alice's laser locks to the injected light, it will adopt the seemingly random phase of that light, which Eve has full knowledge of. If Alice tries to monitor the phase of her light, she will not detect any signs of nonrandomness, even when the attack is successful. The phase-randomized LSA is more difficult to detect, and therefore is the one we implement and analyze in this work. 

Because of the phase-randomized version of the LSA, Alice is forced to rely on optical isolation, or detect the incoming light from Eve using a ``watchdog'' detector, to prevent the attack. It should be noted that neither optical isolation nor watchdog detectors are foolproof solutions: optical isolation can be damaged, and detectors can be blinded, for example \cite{lydersen_hacking_2010, huang_laser-damage_2020}. 

\begin{figure}
	\includegraphics[width=\linewidth]{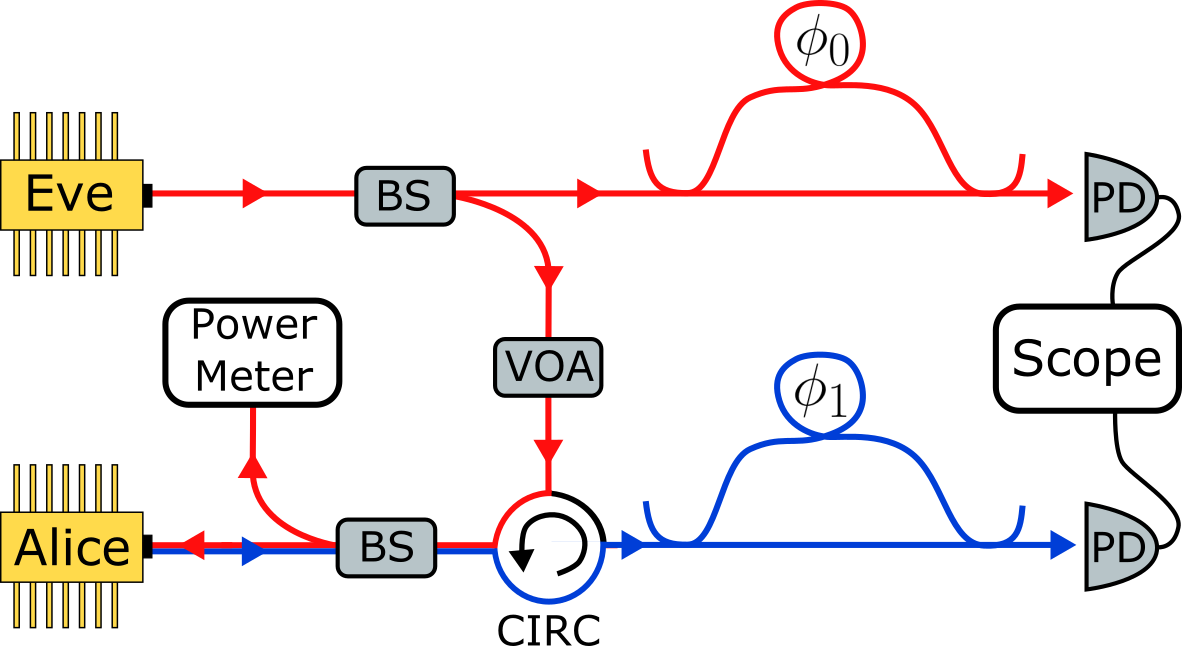}
	\caption{\label{fig:LSA_correlations_setup} Experimental setup for quantifying the phase correlations between Alice's laser and Eve's laser due to the LSA. Using two asymmetric Mach-Zehnder interferometers we can simultaneously monitor the output of Alice's laser (blue lines) and Eve's laser (red lines). The interferometers in effect convert the specific phase of each laser into a specific intensity at the interferometer output, allowing us to monitor the phase of each laser and calculate their correlation. A power meter is used to measure the injected power reaching Alice's unisolated laser. BS, beam splitter; PD, photodiode; VOA, variable optical attenuator.}
\end{figure}

Our experimental setup is shown in Fig.~\ref{fig:LSA_correlations_setup}, where both Alice and Eve monitor the phase of their lasers by using asymmetric Mach-Zehnder interferometers.
In the following we demonstrate how the naive countermeasure of monitoring Alice's phase fails to detect Eve's attack, even when the attack is highly successful. Further, we use this experimental setup to quantify the degree to which Eve is successful in locking Alice's laser in phase as a function of the level of injected optical power. This allows us to, in turn, quantify the level of optical isolation needed to mitigate the attack, in Section \ref{sec:countermeasures}.

Note that some previous experimental demonstrations of the LSA used of isolated laser diodes, requiring much higher injected power to compensate for the isolation losses. More importantly, it is difficult to accurately characterize the isolation of an isolated laser, and therefore it is unclear exactly how much power actually reaches the laser cavity. In our work, we use an unisolated laser (for Alice) and a power meter to accurately measure the injected power that reaches the laser cavity. Therefore, in this paper we use ``injected power'' to mean ``injected power that reaches Alice's laser cavity''.

We use two asymmetric interferometers to monitor the phase of both Eve's and Alice's laser. The intensity at the output of an asymmetric Mach-Zehnder interferometer is given by
\begin{equation}\label{eq:interferometer_intensity_LSA}
	I_{\mathrm{out}} = \frac{I_{\mathrm{in}}}{2}[1 + \cos(\Delta\phi + \phi_0)]
\end{equation}
where $I_\textrm{in}$ is the input intensity, $\Delta \phi$ is the phase difference between the interfering pulses and $\phi_0$ is the relative phase between both arms of the interferometer due to the difference in their length.
We can therefore monitor the intensity at the interferometer output as a measure of the phase difference between successive pulses at the laser output. This allows us to measure the correlation between the phase of both lasers; if the output intensity of Eve's interferometer is highly correlated with that of Alice's, then this indicates a successful attack, giving Eve information about the phase of Alice's laser.

In our experiment, we use nonstabilized interferometers, and therefore $\phi_0$ drifts over time, destroying the correlation between the intensity of both lasers, even if they are locked in phase. To get around this issue, we make many measurements of the correlation, and keep only the maximum and minimum correlation values obtained, which correspond to the times when, by chance, the interferometers are in phase and are out of phase by $\pi$ radians, respectively. Note that a negative correlation coefficient represents an anticorrelation, which is equally beneficial to Eve as a positive correlation. All that matters to Eve is the absolute value of the correlations. In the following, however, we will continue using maximum and minimum correlation values since this most accurately represents our experimental approach. Appendix \ref{sec:appendix_methods} provides further details on this method of measuring the phase correlation even when non-phase-stabilized interferometers are used. 

To set up the experiment, we gain-switch both lasers (distributed feedback telecom laser diodes, unisolated for Alice and isolated for Eve) at 1 GHz using a 3.35 GHz pulse-pattern generator. Our lasers are high-bandwidth (approximately 10 GHz), with a linewidth of approximately 1 MHz, and a tunable wavelength range of  approximately 30 nm, centered at around 1550 nm. Eve's isolated laser has optical isolation of approximately 35 dB. Both lasers are coupled to polarization-maintaining single-mode fiber and are temperature controlled with use of an inbuilt thermoelectric cooler controller. 

We choose 1 GHz as it is representative of the modulation frequencies used in real-world QKD systems, and allows us to reliably generate fully phase randomized pulses. We set the current parameters (duty cycle, bias, and modulation current) appropriately to ensure the phase of each generated pulse is random, and independent of that of prior pulses as previously assessed in Ref.~\cite{lovic_characterizing_2021}. We verify this for both lasers by measuring the autocorrelation of the intensity at the output of each interferometer. Both lasers are disconnected for this measurement, so none of Eve's laser light reaches Alice. 

We then connect both lasers using the circulator and use a variable optical attenuator to adjust the level of injected power. We use polarization maintaining components and fiber throughout, and a manual fiber polarization controller, to match the polarization of both lasers and maximize the effectiveness of the LSA. We measure the average injected power reaching Alice's laser using a power meter. We precisely match the frequencies of both lasers by temperature turning using inbuilt thermoelectric cooler controllers and an optical spectrum analyzer. 

Finally, we measure the intensity waveforms at the output of both interferometers for 25 $\upmu$s, corresponding to 25000 pulses, using two high-speed photodiodes connected to two channels of a high-bandwidth (13 GHz) oscilloscope. 

\begin{figure}
	\includegraphics[width=\linewidth]{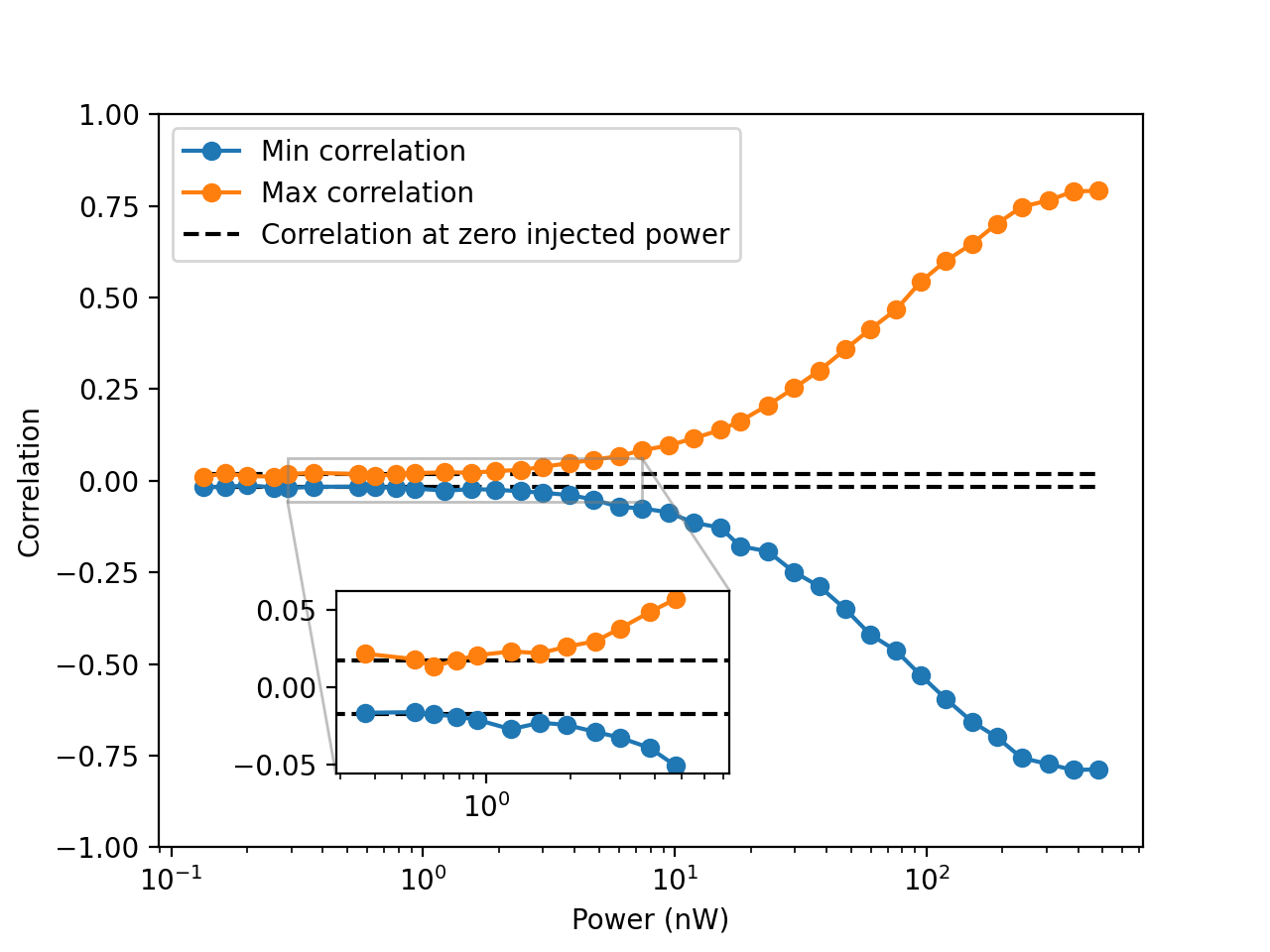}
	\caption{\label{fig:correlations_vs_injected_power} The measured correlation between Alice's laser and Eve's laser as a function of the injected power. Since our interferometers are not phase stabilized, the measured correlation between both lasers drifts along with the interferometer phase. We perform 50 measurements of the correlations at each level of injected power and plot the maximum and minimum values obtained. The dashed black lines indicate the maximum and minimum measured correlations when Eve's laser is disconnected from Alice's laser. Since we plot the maximum and minimum values obtained over 50 measurements, rather than the mean, these lines are not zero due to noise and statistical fluctuations. Any correlations induced by Eve's attack will cause the measured correlations to deviate from this baseline. The inset shows an enlargement the low-power region of the graph, showing that Eve can induce measurable correlations with as little as 1 nW of injected power, which is the point at which the measured correlations start to increase above the baseline.}
\end{figure}

In the following we are particularly interested in the amount of injected power at which Eve starts to induce measurable correlations between her laser and Alice's laser. However, since we are measuring the maximum and minimum values of correlation, rather than the mean, the measured correlations will never be zero, due to statistical fluctuations, even when no light is injected into Alice's laser. Therefore, at zero injected power there will be a minimum and maximum baseline correlation, and we are interested in the amount of injected power at which the correlation rises above this baseline. The measured maximum and minimum baseline correlations at zero injected power will increase as a function of the length of the measurement: the longer a measurement, the higher the likelihood of an outlier correlation measurement pushing the maximum higher or the minimum lower. For this reason, it is important that the length of measurements throughout the experiment remains constant (in our case 25 $\upmu$s). To measure the baseline, we simply disconnect Eve's laser from the circulator before making the correlation measurements (described below). The measured baseline correlation is shown by the dashed black  lines in Fig.~\ref{fig:correlations_vs_injected_power}. Any increase above the maximum correlation baseline must be due to Eve's injected light, and likewise for the minimum correlation. 

To measure the maximum and minimum values of correlation, using a non-phase-stabilized interferometer, we simply repeatedly measure the correlation at different times. 
We make 50 measurements of the correlations at each level of injected power, adjusting the injected power by changing the attenuation of the variable optical attenuator, leaving several seconds between measurements to allow the interferometer phase $\phi_0$ to drift. Our results are plotted in Fig.~\ref{fig:correlations_vs_injected_power}, showing the maximum and minimum values of correlation at each level of injected power. We can clearly see that at high levels of injected power, the LSA is very successful at locking Alice's laser in phase, with correlations between Alice's laser and Eve's laser reaching absolute values above 0.8. In practice, the maximum and minimum correlation values will never reach $1$ and $-1$, since other sources of noise, such as intensity noise, or chirp, will also influence the interferometer output, acting against the locking effects of the LSA and reducing the correlation. 

Therefore, for injected powers above 100 nW, Eve has near perfect knowledge of the phase of Alice's emitted pulses. From her perspective, the pulses are almost fully nonrandom, which undermines the security of the communications due to the drastically lower secure key rates of non-phase-randomized QKD. 

\begin{figure}
	\includegraphics[width=\linewidth]{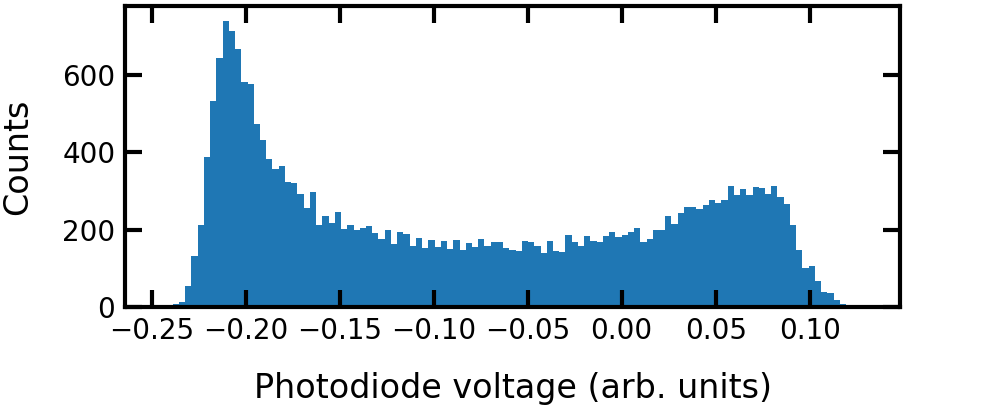}
	\includegraphics[width=\linewidth]{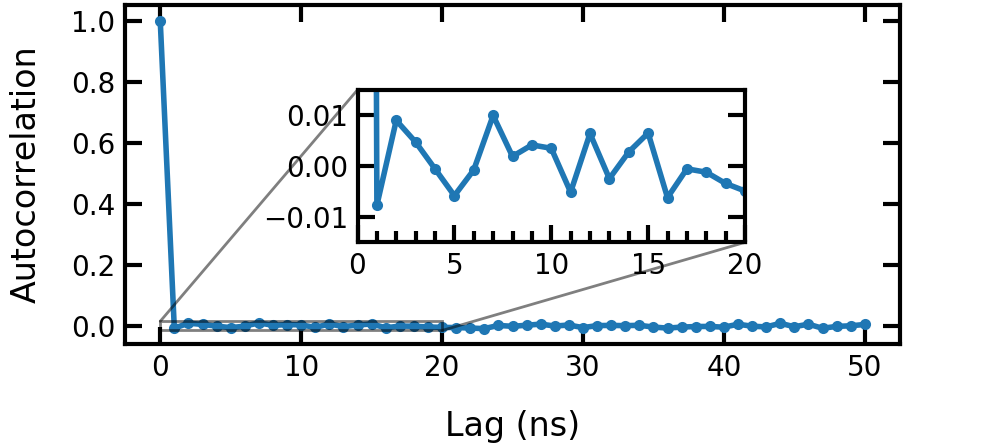}
	\caption{\label{fig:distribution_and_autocorrelation} The distribution (top) and autocorrelation (bottom) of the intensity at the output of Alice's interferometer when Eve is injecting 480 $\upmu W$ of optical power. This corresponds to the highest power measurement in Fig.~\ref{fig:correlations_vs_injected_power}, with a correlation between Alice's laser and Eve's laser above 0.8. Despite Eve obtaining near perfect information about the phase of Alice's laser, Alice cannot detect this by monitoring the output of her laser, since both the distribution and the autocorrelations are indicative of a fully phase-randomized laser. The distribution (top) is flatter and more asymmetric than an ideal arcsine distribution due to sources of noise such as jitter, chirp, and intensity noise, which are significant when the laser is modulated at high frequencies as is common in QKD.}
\end{figure}

Note that despite the large correlation between both lasers, the output of Alice's laser does not display any signs of nonrandomness. We plot the distribution of intensity at the output of Alice's laser in Fig.~\ref{fig:distribution_and_autocorrelation} (top), showing the signature arcsine distribution indicative of full phase randomization. Additionally, we plot the autocorrelation of the intensity in Fig.~\ref{fig:distribution_and_autocorrelation} (bottom), again showing no signs of nonrandomness. This demonstrates that if Eve uses phase randomized light to perform the LSA, Alice cannot detect the attack by monitoring the output of her laser. Instead, to prevent the attack, Alice is forced to either block the incoming light using optical isolation or detect it with a monitor ``watchdog'' detector.

As the injected power decreases, so do the correlations. The inset in Fig.~\ref{fig:correlations_vs_injected_power} shows an enlargement of the lower-power measurement results, showing that, in our experiment, Eve can induce measurable correlations with as little as 1 nW of injected power. Starting at around 1 nW, the measured correlations rise above the zero injected power baseline, showing that Eve's attack increases the correlation between both lasers. We stress that this 1 nW threshold is specific to our experiment. Other experiments, and indeed practical QKD systems, will undoubtedly have different thresholds, depending on the lasers being used, experimental conditions such as temperature, and other factors. However, our observations and analysis are widely applicable to all QKD transmitters under the LSA. 

One nanowatt is a much lower level of injected power than used in previous studies analyzing the LSA, and suggests that the LSA requires a lower level of injected power to be successful than previously thought \cite{sun_effect_2015, huang_laser-seeding_2019}. However, note that 1 nW is the threshold at which the effects of the LSA just start to become measurable, and are therefore still very small. Indeed, our results are still consistent with those of prior studies: at 100 nW of injected power, similar to the injected power used in the work reported in Ref.~\cite{huang_laser-seeding_2019}, we observe very large effects from the LSA consistent with the results reported there. 

At this point, it is worth recalling that for the standard decoy protocol, as pointed out in Ref.~\cite{sun_effect_2015}, even a minimal deviation from phase randomization would place the protocol outside the framework of assumptions necessary to guarantee its unconditional security. This conservative standpoint, therefore, prevents us from assuming that the absence of measurable correlations below 1 nW implies perfect phase randomization. In this context, recently introduced security frameworks may prove beneficial in obtaining a secret key rate even with imperfect phase randomization \cite{curras-lorenzo_security_2023, sixto_secret_2023, nahar_imperfect_2023}. In particular, the situation explored here, where Eve can arbitrarily set each pulse phase value, aligns with the case of zero-length correlation between pulses and a nonuniform phase distribution, as discussed in Ref.~\cite{curras-lorenzo_security_2023}.

Nonetheless, it is plausible that a situation in which correlations are undetectable corresponds to a weaker coupling between Alice's laser and Eve's laser. As a result, this can still be considered a favorable initial condition to mitigate the LSA. To this end is worth investigating the level of isolation necessary to prevent levels higher than 1 nW reaching Alice’s laser cavity. Using this minimum injected power threshold, in the next section, we take an approach similar to that in Ref.~\cite{lucamarini_practical_2015} to quantify the optical isolation required to prevent a successful LSA.

\subsection{Countermeasures}\label{sec:countermeasures}
To prevent the LSA and guarantee the security of the QKD communications, the transmitter must implement countermeasures. This could consist of a ``watchdog" detector placed at the entrance of the QKD transmitter, monitoring any incoming light \cite{lucamarini_practical_2015}. If this detector registers incoming light above some threshold value, then the communications can be aborted. However, this solution comes with several drawbacks. Most obvious is the increased cost and complexity of the QKD system. Another disadvantage is that there are several known security vulnerabilities of detectors used in QKD systems \cite{xu_secure_2020}. For example it has been shown that certain types of detector can be ``blinded" by the shining of bright light on them, making them unresponsive \cite{lydersen_hacking_2010}. In general, active countermeasures, as opposed to passive ones, are less desirable because of their increased complexity, which can introduce further avenues to attack, such as the aforementioned blinding. Instead, passive countermeasures are preferable. In particular, Alice can use optical isolation to block any incoming light \cite{ponosova_protecting_2022}. An optical isolator is a device that is transparent to light traveling in one direction but opaque to light traveling in the other direction. Clearly, an ideal optical isolator would be a suitable countermeasure. However, optical isolators are not ideal, and in particular they are not able to completely block light, but rather they just attenuate the light by a large amount. Therefore, Alice cannot completely prevent light from entering her transmitter, but can only control the attenuation of the incoming light. 

A key quantity is therefore the minimum amount of light that needs to reach Alice's laser for Eve's attack to be successful. Clearly, if any amount of injected light, however small, will lead to a successful attack, then no amount of optical isolation can prevent it. In the previous section, we showed how we experimentally determined that, for our experimental apparatus and conditions, Eve can induce measurable correlations between the phase of her laser and that of Alice's laser by injecting as little as 1 nW of (average) optical power. In the following, we use 1 nW as an example minimum power threshold for a successful LSA, but we stress that other experiments and QKD systems may well have different thresholds, which should be verified independently. A second key quantity is the maximum amount of light that Eve can inject into Alice's transmitter. Clearly, if Eve can inject an unlimited amount of light, then no amount of optical isolation can guarantee the 1 nW limit. 

One upper limit to the amount of light that Eve can inject is given by the laser-induced damage threshold (LIDT). The LIDT is defined as the maximum power that can be transmitted through an optical fiber without damaging it. Lucamarini \textsl{et al.} Ref.~\cite{lucamarini_practical_2015} quote a value for the LIDT of standard single-mode silica optical fiber of 55 kW, beyond which the fiber softens and begins to melt. Reducing 55 kW of optical power to 1 nW would require on the order of 140 dB of optical isolation. Alternatively, we can upper bound the amount of light Eve can inject by using an optical fuse, which is a device intended to break and stop the transmission of any light if the power goes above some threshold.  An optical fuse with a threshold of a few watts, reducing the required optical isolation to approximately 90 dB, was introduced in Ref.~\cite{shin-ichi_optical_2004}. There also exist so-called optical power limiters, which do not break above a certain threshold like an optical fuse, but rather prevent the transmitted optical power from increasing beyond that threshold. A recent proposal introduced a simple power limiting device with a widely tunable power threshold, down to values of around 1 $\upmu$W, which would reduce the optical isolation requirement down to approximately 30 dB \cite{zhang_securing_2021}.

\begin{table}
	\caption{\label{tab:rate_equation_parameters} Rate equation parameters used in the simulations. }
	\resizebox{\linewidth}{!}{%
		\begin{tabular}{ l c l } \toprule
			
			Parameter & Value & Description \\ \midrule
			
			$\tau_n$ (ns) & 0.15 & Carrier lifetime\\
			$\tau_p$ (ps) & 4.47 & Photon lifetime \\
			$g$ ($\textrm{cm}^3\textrm{s}^{-1}$) & 1.70 $\times 10^{-6}$ & Differential gain coefficient \\
			$\varepsilon$ ($\textrm{cm}^3$) &$ 3.24 \times 10^{-17}$& Gain compression factor \\
			$N_0$ ($\textrm{cm}^{-3}$) & $3.79 \times 10^{18}$ & Carrier density at transparency \\
			$\beta$ & $4.44 \times 10^{-5}$ & Spontaneous emission factor\\
			$\alpha$ & 2.95 & Linewidth enhancement factor\\
			$\eta$ & 0.52 & Differential quantum efficiency\\
			$V$ ($\textrm{cm}^3$) & $2\times10^{-11}$ & Active layer volume\\
			$\Gamma$ & 0.22 & Mode confinement factor \\
			$\kappa$ (Hz) & $1.13 \times 10^{11}$& OIL coupling term \\ \bottomrule
			
	\end{tabular}}
\end{table}

\section{Numerical Model of the LSA}\label{sec:LSA_OIL}
Although the effect of the LSA on the phase randomization of a gain-switched laser is perhaps the most damaging, the LSA has many other effects on the laser output and Eve can try to use any of these to her advantage. In this section we model the effects of the LSA in simulation, by considering it as an instance of  (OIL) \cite{liu_optical_2020}. We present the OIL laser rate equations which can serve as a general model of the LSA and can be used to study all of the effects of the LSA, including the phase randomization, but also other effects, such as an increase in the power output, a reduction in the turn-on delay, or a reduction in the chirp and jitter.

The OIL rate equations consist of two coupled sets of differential equations, one for modeling the master laser and one for modeling the slave laser \cite{paraiso_advanced_2021}. Eve's master laser is described by the standard rate equations for a free-running semiconductor laser that describe the rate of change of the carrier number ($N$), the photon number ($S$) and the phase ($\phi$) \cite{cartledge_extraction_1997, fatadin_numerical_2006}:
\begin{align}
	\frac{d N(t)}{d t} &=\frac{I(t)}{qV}-\frac{N(t)}{\tau_{n}}-g \frac{N(t)-N_{0}}{1+\epsilon S(t)} S(t) + F_N(t) \label{eq:N} \\
	\frac{d S(t)}{d t} &=\Gamma g \frac{N(t)-N_{0}}{1+\epsilon S(t)} S(t)-\frac{S(t)}{\tau_{p}}+\frac{\Gamma\beta N(t)}{\tau_{n}} + F_S(t) \label{eq:S} \\
	\frac{d \phi(t)}{d t} &=\frac{\alpha}{2} \left[ \Gamma g(N(t)-N_0) -\frac{1}{\tau_p}\right] + F_\phi(t) \label{eq:phi}
\end{align}
where $I(t)$ is the applied current, $q$ is the electron charge, $V$ is the active layer volume. $\tau_n$ and $\tau_p$ are the carrier and photon lifetimes, respectively, which quantify the average time a carrier or photon survives in the laser cavity. $\Gamma$ is the mode confinement factor, which accounts for the fact that only a fraction $\Gamma$ of the photons are confined to the active layer, $g$ is the differential gain coefficient which arises from our making the approximation that the gain is linear as a function of carrier density, $\epsilon$ is the gain compression factor, which accounts for the non-linear reduction in gain at high power outputs \cite{koch_effect_1986}, $N_0$ is the carrier density at transparency, $\beta$ is the fraction of spontaneous emission coupled into the lasing mode, $\alpha$ is the linewidth enhancement factor which quantifies the increase in linewidth due to the coupling between refractive index and carrier density in semiconductor lasers \cite{henry_theory_1982}, and $F_{N}$, $F_{S}$, $F_{\phi}$ are Langevin noise terms, which capture the effects of spontaneous emission. These terms are defined in Appendix \ref{sec:appendix_rate_equations}. The power output of the laser is related to the photon density by
\begin{equation} \label{eq:power}
	P(t)=\frac{V \eta h \nu}{2\Gamma \tau_{p}} S(t)
\end{equation}
where $\eta$ is the differential quantum efficiency, $h$ is Planck's constant, and $\nu$ is the laser frequency.

Alice's slave laser cavity receives additional photons from Eve's master laser and, to account for these injected photons, the standard free-running rate equations need to be extended as follows \cite{troger_novel_1999, lau_enhanced_2009, liu_optical_2020}:
\begin{align}
\frac{dN(t)}{dt} &= \frac{dN_\textrm{fr}(t)}{dt} \label{eq:N_OIL} \\
\frac{dS(t)}{dt} &= \frac{dS_\textrm{fr}(t)}{dt} +2 \kappa \sqrt{S_{\text {inj }}(t) S(t)} \cos(\Delta\phi(t) - \Delta\omega\inj t) \label{eq:S_OIL} \\
\frac{d\phi(t)}{dt} &= \frac{d\phi_\textrm{fr}(t)}{dt} -\kappa \sqrt{\frac{S_{\mathrm{inj}}(t)}{S(t)}} \sin (\Delta\phi(t) -\Delta \omega\inj t)  \label{eq:phi_OIL}
\end{align}
where the subscript fr denotes the standard rate equations for a free running laser given by Eqns. \ref{eq:N}-\ref{eq:phi}, $\Delta\phi = \phi(t) - \phi\inj(t)$ is the difference between the secondary laser phase and the phase of the injected light,
$\kappa$ is a coupling coefficient that quantifies the rate at which injected photons enter the secondary laser cavity, $S\inj$ is the injected photon density and $\Delta \omega\inj$ is the difference in free-running optical angular frequency between primary laser and secondary laser. In the following simulations, we use rate equation parameters obtained by the fitting of parameters to experimental measurements of a distributed feedback laser \cite{bjerkan_measurement_1996, cartledge_extraction_1997}, given in Table \ref{tab:rate_equation_parameters}. 

\begin{figure}
	\includegraphics[width=\linewidth]{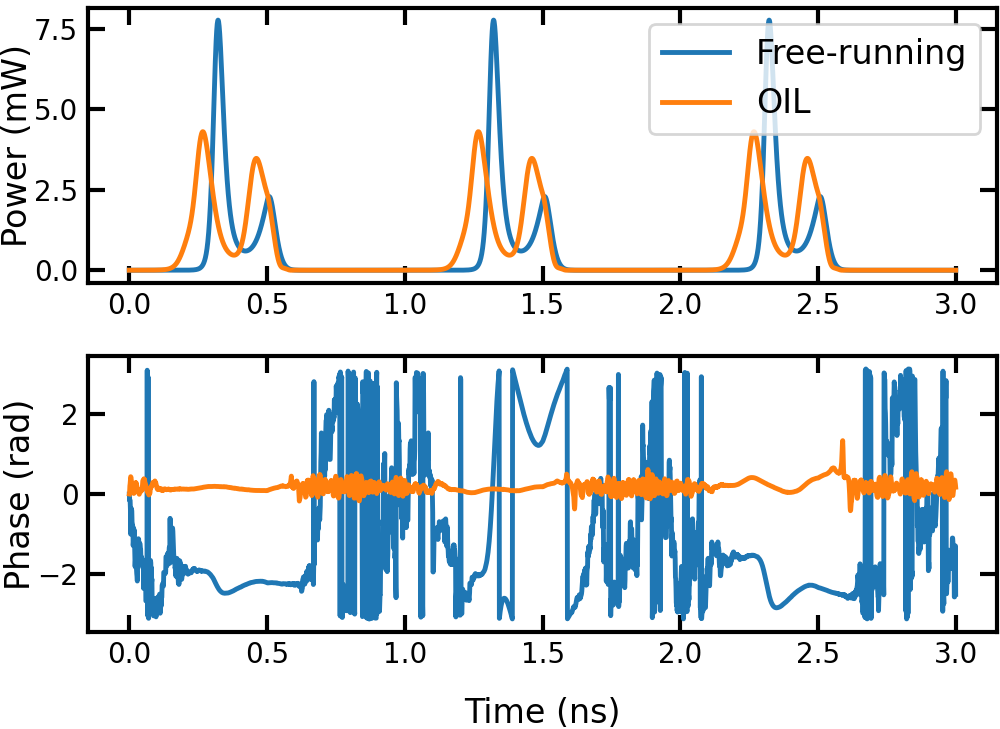}
	\caption{\label{fig:pulse_shape_and_phase}A rate equation simulation of the power output (top) and phase (bottom) of a gain-switched laser with and without the effects of OIL. The top plot shows that with the parameters in this study, OIL causes the laser to start lasing sooner and have a different pulse shape, leading to an overall increase in total energy emitted. The bottom plot shows that the free-running laser experiences drastic phase diffusion while the laser is below threshold and not lasing. OIL suppresses this phase diffusion and keeps the phase at around a constant value set by the master laser.}
\end{figure}
Using the rate equations, we can simulate the power output and phase of Alice's gain-switched slave laser, both with and without the effects of the LSA. The blue curve in Fig.~\ref{fig:pulse_shape_and_phase} plots the power output and phase of a free-running gain-switched laser, and is obtained by our solving the free-running laser rate equations without OIL terms. It displays signature features of gain-switching: a train of pulses, each pulse with one or more relaxation oscillation peaks, and each with a random phase, due to rapid phase randomization between pulses. The orange curve in Fig.~\ref{fig:pulse_shape_and_phase} plots the power output and phase of a gain-switched laser under the effects of the LSA. It is obtained by our solving the rate equations with OIL terms and setting $P\inj = 100$ nW and $\phi\inj = 0$. These two parameters can be set to any values to simulate the effects of the LSA under different levels of injected optical power and different phase differences between the master laser and the slave laser.  $P\inj$ and $\phi\inj$ do not even need to be constant if, for example, Eve's master laser is itself gain-switched. 

\begin{figure}
	\includegraphics[width=\linewidth]{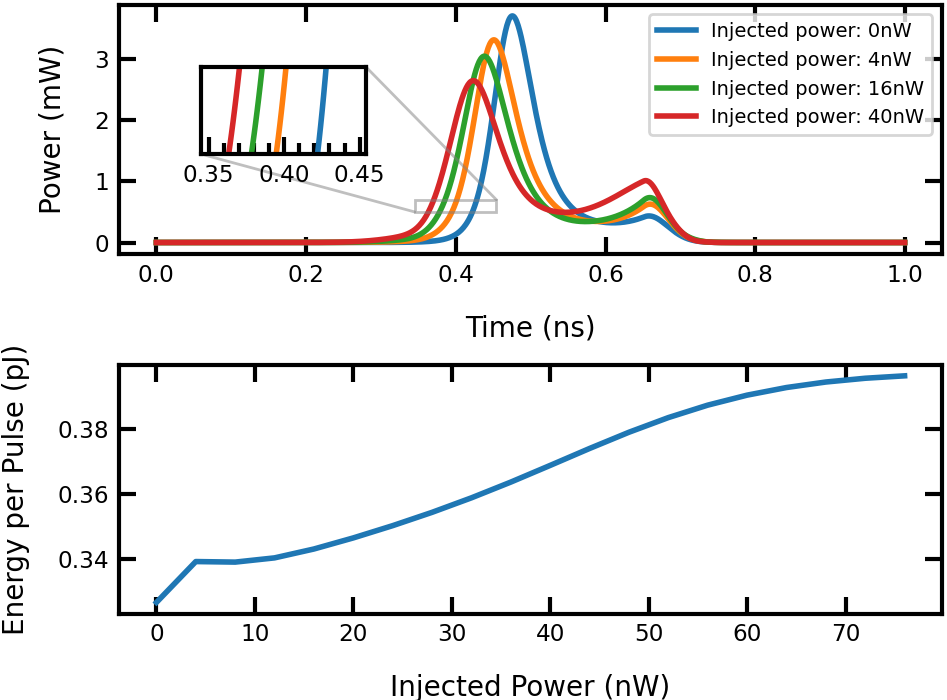}
	\caption{\label{fig:energy_vs_injected_power} The change in power output of the laser under various levels of injected power (top). The pulse shape changes with the injected power, and in particular there is a shorter turn-on delay (the laser starts lasing sooner; see the inset), leading to an increase in total energy emitted per pulse which can be calculated by integration of the pulse over one period.  The energy per pulse increases as a function of injected power (bottom). This simulation result is in good agreement with previous experimental measurements \cite{huang_laser-seeding_2019}.}
\end{figure}
The orange curve in Fig.~\ref{fig:pulse_shape_and_phase} displays several signature effects of the LSA on the output of a gain-switched laser. The effect on the phase randomization is clear:  the LSA prevents the slave laser phase from being randomized, keeping it centered at around a constant value set by the phase of the master laser $\phi\inj$ (in this case 0). This is in agreement with the experimental results demonstrating the LSA in Ref.~\cite{sun_effect_2015}.  Likewise, the effect on the power output is apparent: the slave laser begins lasing sooner (shorter turn-on delay), and the relaxation oscillation peaks are shorter and wider. It was demonstrated in Ref.~\cite{huang_laser-seeding_2019}  that an increase in the photon number per pulse, or equivalently energy per pulse, emitted by a QKD transmitter is detrimental to its performance and security. We can calculate the energy per pulse in simulation by integrating the simulated laser power output over one period. 

To investigate how the energy per pulse changes as a function of injected optical power, we repeat the simulations under various levels of injected power $P\inj$. Fig.~\ref{fig:energy_vs_injected_power} (a) shows the resulting pulse shapes, indicating that at higher levels of injected power, the slave laser begins lasing sooner (reduced turn-on delay, indicated by the inset), while the first relaxation oscillation peak reduces in power and the second peak increases in power. By integrating these curves over one period, we can plot the energy per pulse as a function of injected power, and Fig.~\ref{fig:energy_vs_injected_power} (b) shows an increase in energy with higher levels of injected power. These simulations again agree with previous experimental results \cite{huang_laser-seeding_2019}, and demonstrate the usefulness of the rate equation model in studying the LSA. We have shown that the rate equation model can simulate the main effects of the LSA studied experimentally to date.

 In the next section we demonstrate how the rate equation model can go further and can be used to generate new insight into the LSA. We study how the effects of the LSA on the total energy emitted per pulse changes when using different laser current parameters, and we verify our numerical simulations with experimental measurements. We argue that a rate equation model can be a useful tool for exploring large ranges of parameters, which would be time-consuming to explore experimentally.

\section{New insight into the LSA using a rate equation model} \label{sec:intensity}
\begin{figure}
	\includegraphics[width=\linewidth]{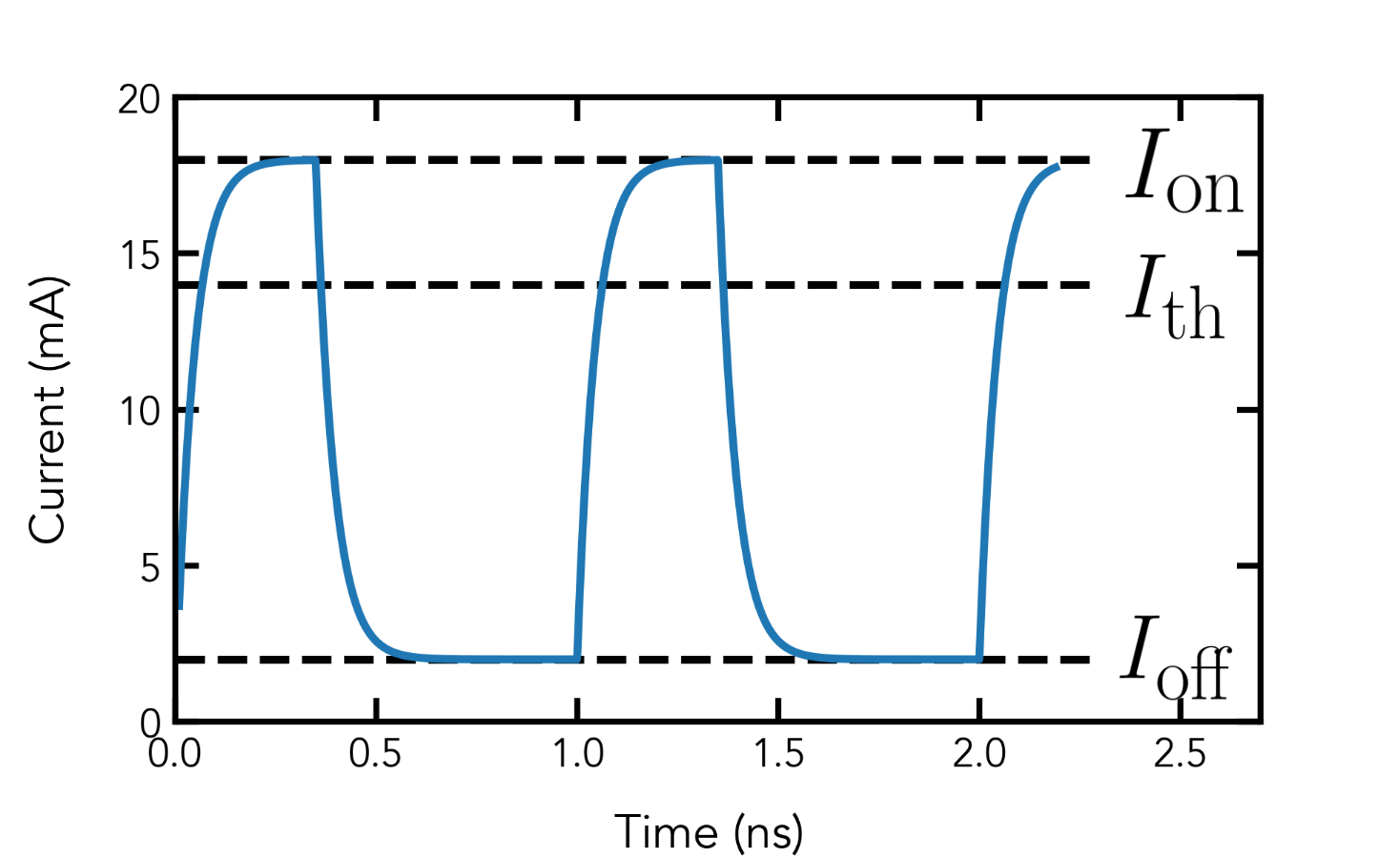}
	\caption{\label{fig:current}The laser driving current alternates between an on-time current $I_\textrm{on}$ and an off-time current $I_\textrm{off}$. The duty cycle is calculated to achieve single-peak emission, by our setting the duration of the current on-time to equal the laser turn-on delay plus half the period of a relaxation oscillation. To account for the finite bandwidth of the driving electronics, a 3.5 GHz low-pass filter is applied to the current.}
\end{figure}

The pulses emitted by a QKD transmitter need to be short, phase randomized and with a photon number close to zero \cite{yuan_interference_2014}. The last requirement can be met with an arbitrary level of attenuation after the laser. The first two requirements can instead be obtained by suitable adjustment of the current signal that drives the gain-switched laser. Therefore, we can ask whether certain current driving parameters are more susceptible to the LSA than others, in terms of energy increase. We can easily investigate this question using the rate equation model by simply adjusting the current parameter $I(t)$ and calculating the energy increase. In this way, we can explore hundreds of parameters in simulation, before undertaking time-consuming experimental work.

We can define the injected current as a 1 GHz pulse wave, with an on-time current  $I_\textrm{on}$ and an off-time current  $I_\textrm{off}$ and a variable duty cycle. We choose to define the current in terms of on-time and off-time current, instead of the more typical bias and modulation current, because this makes it easier to define constraints such that, for example, the current is never negative. 

To model the finite bandwidth of the pulse generator used to drive the laser, we apply a 3.5 GHz low-pass filter to the injected current. Fig.~\ref{fig:current} gives a visual representation of the driving current. 

We can then vary $I_\textrm{on}$ and $I_\textrm{off}$ across a wide range of values subject to the following constraints: the off-time current of the laser should never be negative, and should always remain below the threshold current, i.e. $0<I_\textrm{off}<I_\textrm{th}$, where $I_\textrm{th}$ is the threshold current; and the on-time current should always be above the threshold current, i.e., $I_\textrm{on}>I_\textrm{th}$. To guarantee the requirement of narrow pulses, we set the duty cycle such that the on-time of the pulse wave is equal to the turn-on delay plus half the period of the relaxation oscillations \cite{shakhovoy_influence_2021}.

\begin{figure}
	\includegraphics[width=\linewidth]{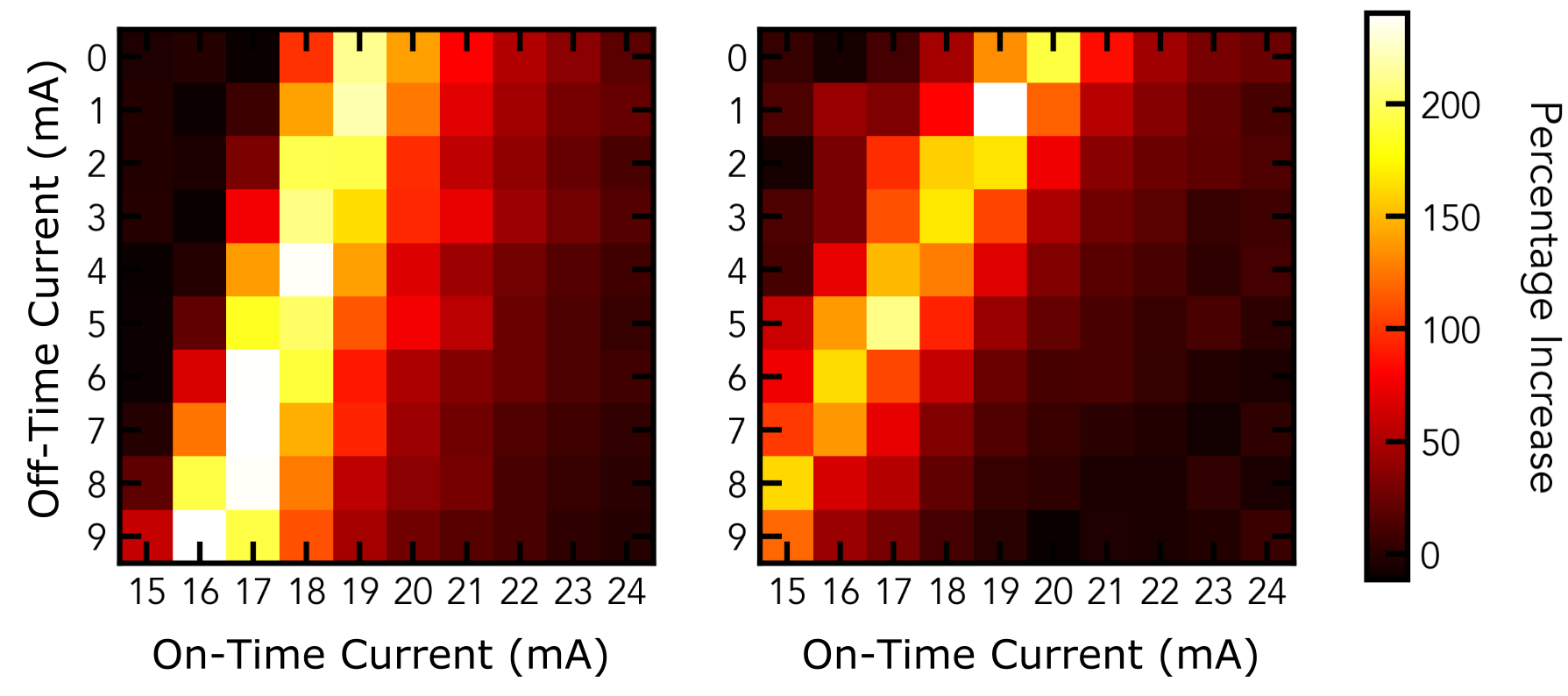}
	\caption{\label{fig:heatmaps}Simulated (left) and experimental (right) heatmaps of the percentage energy per pulse increase of Alice's laser with different current driving parameters. We drive the laser with a range of current parameters, from 0 to 9 mA of off-time current, and from 15 to 24 mA of on-time current. For each combination of current parameters (each square in the heatmaps) we calculate the percentage increase in intensity per pulse caused by the LSA. The correspondence between simulation and experiment is evidence that the rate equation model accurately describes the LSA. These measurements indicate that the effects of the LSA on the power output of a laser is highly dependent on the specific current parameters used to drive the laser.}
\end{figure}

Fig.~\ref{fig:heatmaps} (a)  shows our simulation results in the form of a heatmap. 
For each combination of current parameters $(I_\textrm{on}, I_\textrm{off})$, we calculate the percentage increase in energy due to the LSA by solving the rate equations with and without injected light. To calculate the energy per pulse without injected light, we set $P\inj = 0$ to model a free-running laser, and integrate the solution to the rate equations over one period. To calculate the energy per pulse with injected light, we set $P\inj = 100$ nW, which is a level of injected power used in previous experimental studies of the LSA \cite{huang_laser-seeding_2019, pang_hacking_2020}. The heatmap color bar indicates the calculated percentage increase in energy per pulse. 

Using the same setup as in Fig.~\ref{fig:LSA_correlations_setup}, we can experimentally verify the simulation results shown in Fig.~\ref{fig:heatmaps} (a). For each combination of current parameters covered by the heatmap, we measure the average output power of Alice's laser with no injected light and with 100 nW of continuous wave injected light from Eve's laser. Our experimental results are shown in Fig.~\ref{fig:heatmaps} (b), and agree well with the simulations. In particular, both heatmaps display a notable line of light-colored squares rising diagonally from the bottom left corner, with darker squares elsewhere. 
The good agreement between experiment and simulation is evidence of the accuracy of the rate equation model in describing the LSA. On the basis of these simulations, we can see that the specific current parameters, i.e. the specific square on the heatmap, used to drive a gain-switched laser can have a large impact on the effect of the LSA on the power output of that laser. 

There are many factors to consider when one is choosing the current parameters to drive a gain-switched laser in a QKD transmitter. The shape of the emitted pulses, their width, chirp and other factors are all important. The simulation and experimental results in Fig.~\ref{fig:heatmaps} can serve as another data point and consideration when one is choosing these parameters. For example, Alice may choose to avoid a bright square so as to minimize the energy increase caused by the LSA. On the other hand, a large increase in energy could be used to detect the presence of Eve implementing the LSA on her laser. The heatmaps in Fig.~\ref{fig:heatmaps} can serve as one factor among others when one is choosing the best current parameters for a gain-switched laser in a QKD system. 

These results demonstrate the usefulness of a rate equation model for studying the LSA. Using the model, we were able to explore hundreds of parameters in simulation before conducting time-consuming experimental work. Although we focused on the energy increase, other effects of the LSA can similarly be studied, such as the reduction in the turn-on delay.

\section{Conclusion} \label{sec:conclusion}

In this work we presented an experimental and numerical study of the laser seeding attack in QKD. We implemented the LSA experimentally and measured the effect on the phase randomization as a function of injected optical power. This allowed us to quantify the level of optical isolation required to mitigate the attack. Since the effect on the phase is just one of multiple effects of the LSA, we introduced the laser rate equations as a useful tool to study and model the LSA, drawing on the literature on optical injection locking from the field of classical telecommunications. Using this model, we were able to reproduce previously published experimental results, and demonstrate how the model can be used to study other properties of the LSA. In particular, we measured the increase in power output of a gain-switched laser when subjected to the LSA with different laser current parameters. We found that certain parameters made the laser more vulnerable to the LSA, and we verified our findings experimentally. Altogether, our work contributes to the security of QKD transmitters, and provides a tool for further investigations of the LSA and other attacks that target the laser in a QKD transmitter.

\begin{acknowledgments}
The work reported here was funded by the project EMPIR 19NRM06 METISQ, which received funding from the European Metrology Programme for Innovation and Research (EMPIR) cofinanced by the participating states and from the European Union’s Horizon 2020 research and innovation program.
V. L. acknowledges financial support from the EPSRC (EP/S513635/1) and Toshiba Europe Ltd.
\end{acknowledgments}

\appendix
\section{Measuring phase correlation with unstable interferometers}\label{sec:appendix_methods}
Since both interferometers are not phase stabilized, the relative phase between both arms of the interferometer drifts between 0 and $2\pi$. In terms of Eq. \ref{eq:interferometer_intensity_LSA}, $\phi_0$ is not stabilized and instead drifts between 0 and $2\pi$, at a slow rate on the order of radians per second. This is much slower than our measurement time of 25 $\upmu s$, and therefore has a negligible effect on an individual measurement. However, for measurements taken at different times, $\phi_0$ can take on very different values, requiring care when one is comparing results across measurements taken at different times. For example, if both interferometers have the same value of $\phi_0$, and both lasers are locked in phase ($\Delta\phi = 0$), then the correlation between the waveforms will be 1. On the other hand, if there is a phase difference of $\pi$ between both interferometers then the correlation will be $-1$. And if the phase difference is $\pi/2$, then the measured correlations will be 0, even though both lasers are still locked in phase. Given that the value of $\phi_0$ for both interferometers drifts over time, we should expect to see the correlations correspondingly drift between a maximum value and a minimum value, with a mean value of 0. We stress that this drift of the correlation is due to the unstable interferometers, and is not due to the change in underlying correlation between the phase of Alice's laser and Eve's laser.

\begin{figure}
	\includegraphics[width=\linewidth]{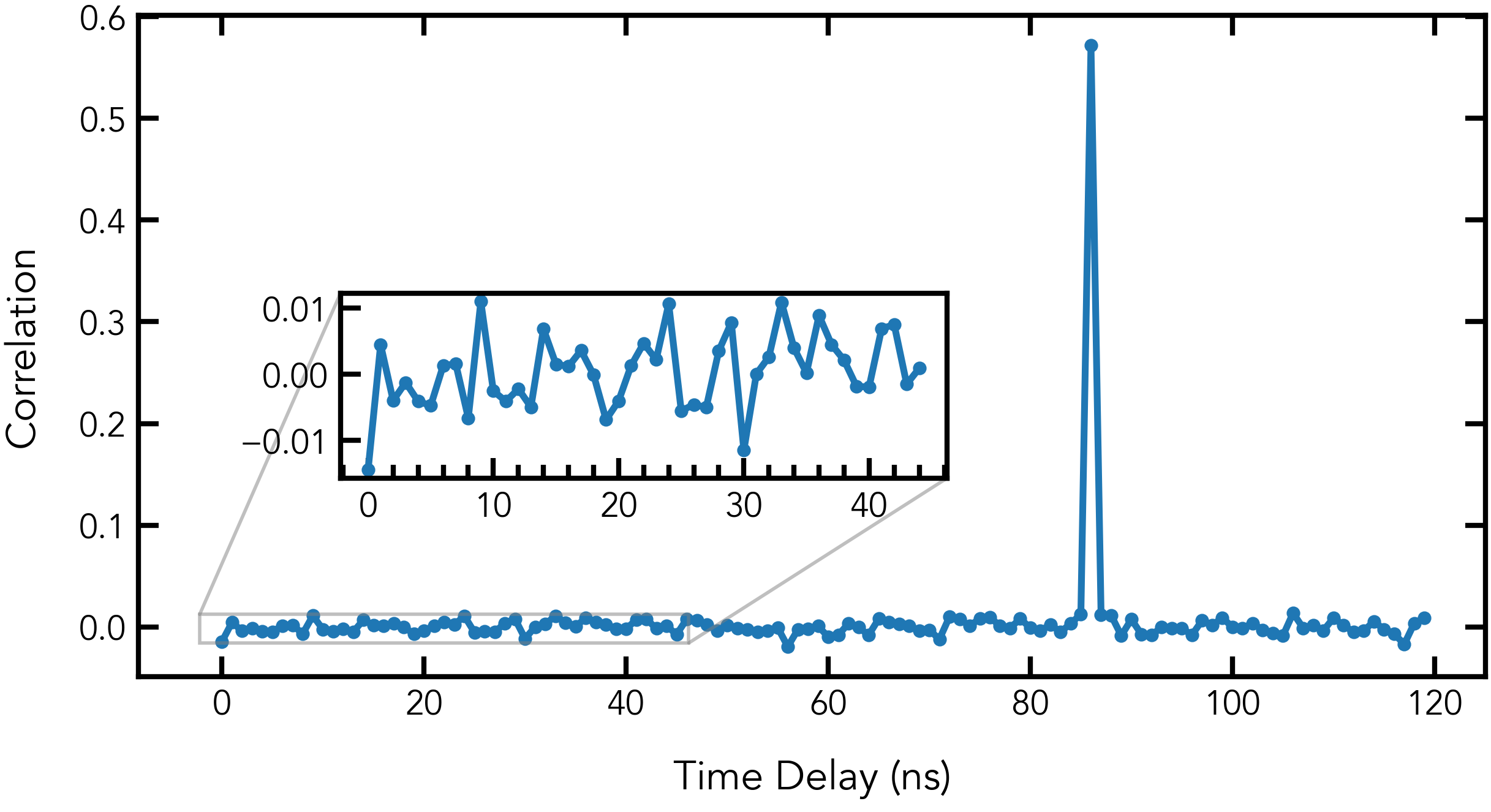}
	\caption{\label{fig:correlations_vs_lag}The correlation between Alice and Eve's lasers as a function of time-delay. Given that it takes some time for Eve's light to reach Alice, the output of Eve's laser will be correlated to the output of Alice's laser with some time delay. We calculate the correlations as a function of time-delay and find a very large correlation between both lasers 86 ns apart.}
\end{figure}
Note that the oscilloscope does not measure the waveforms on both channels at exactly the same time. Furthermore, it takes Eve's light a few nanoseconds to reach Alice's laser, which further adds to the time mismatch between both measured waveforms. To account for this mismatch, we calculate the correlation between both waveforms as a function of time delay, or lag, by shifting one waveform with respect to the other by 1 ns intervals. An example measurement is plotted in Fig.~\ref{fig:correlations_vs_lag}, showing no correlations at most lags, and a very high correlation at a lag of 86 ns. This indicates that the combined effect of the time mismatch between the waveform measurements on both oscilloscope channels and the time delay of Eve's light reaching Alice laser, equals 86 ns. In subsequent measurements, we therefore calculate the correlations between both waveforms displaced by 86 ns.

\section{Laser Rate Equation Noise Terms}\label{sec:appendix_rate_equations}
To account for spontaneous emission noise, additional terms are added to the laser rate equations. These so-called Langevin noise terms take on different forms depending on the sources of noise being considered. To account for the effects of spontaneous emission, they are given by: 

\begin{align}
	&F_{S}(t)=\sqrt{\frac{2 \Gamma \beta N(t) S(t)}{\tau_{n} \Delta t}}  x_{S} \\
	&F_{\phi}(t)=\sqrt{\frac{\Gamma \beta N(t)}{2 \tau_{n} S(t) \Delta t}}  x_{\phi} \\
	&F_{Z}(t)=\sqrt{\frac{2 N(t)}{V \tau_{n} \Delta t}  x_{Z}} \\
	&F_{N}(t)=F_{Z}(t)-\frac{F_{S}(t)}{\Gamma}
\end{align}

where $F_Z(t)$ is a noise term, uncorrelated to $F_S(t)$ and $F_\phi(t)$, used to define the carrier density noise term $F_N(t)$. $\Delta t$ is the integration time step and $x_S$, $x_\phi$, and $x_Z$ are three independent standard normal random variables. Often the rate equations are used without noise terms, when the effects of noise are not of interest. In this case, the rate equations can be solved with use of standard numerical integration tools. However, when the noise terms are included, the rate equations become stochastic differential equations and must be solved using by stochastic numerical integration methods,the simplest of which is the Euler–Maruyama method \cite{higham_algorithmic_2001}. 

\bibliographystyle{ieeetr}

\end{document}